\documentclass[AMA,Times1COL]{WileyNJDv5} 

\usepackage{graphicx}
\articletype{Article Type}%

\received{Date Month Year}
\revised{Date Month Year}
\accepted{Date Month Year}
\journal{Journal}
\volume{00}
\copyyear{2025}
\startpage{1}

\begin{document}

\title{Test-Negative Designs with Multiple Testing Sources}

\author[1]{Mengxin Yu}

\author[2]{Nicholas P. Jewell}

\authormark{Yu \textsc{et al.}}
\titlemark{Test-Negative Designs with Multiple Testing Sources}

\address[1]{\orgdiv{Department of Statistics and Data Science}, \orgname{University of Pennsylvania}, \orgaddress{\state{PA}, \country{United States}}}

\address[2]{\orgdiv{Department of Medical Statistics}, \orgname{ London School of Hygiene \& Tropical Medicine}, \orgaddress{\state{London}, \country{United Kingdom}}}

\corres{Corresponding author: Nicholas P. Jewell, \email{nicholas.jewell@lshtm.ac.uk}}


\abstract[Abstract]{Test-negative designs, a form of case-cohort studies, have been commonly used to assess infectious disease interventions. Early examples of the design included the evaluation of seasonal influenza vaccines in the field. Recently, they have also been widely used to evaluate the efficacy of COVID-19 vaccines in preventing symptomatic disease for different variants \citep{evansjewell2021}. The design hinges on individuals being tested for the disease of interest; upon recruitment, such individuals are subjected to a definitive test for the presence of the disease of interest (test-positives), or not (test-negatives) along with the determination of whether the individual has been exposed to the intervention under study (e.g. vaccination). In most early TND studies, individuals were tested because they were suffering from symptoms consistent with the disease in question, and the TND was a tool to reduce confounding due to healthcare-seeking behavior. However, in many cases, such as COVID-19, and Ebola, testing results were available at healthcare facilities for individuals who presented for a variety of reasons in addition to symptoms (e.g., case contact tracing, etc.). Aggregating samples from symptomatic and asymptomatic test results leads to bias in the assessment of the efficacy of the intervention. Here we consider these issues in the context of a specific version of the `\emph{multiple reasons for testing problem},' motivated by a vaccine trial designed to assess a new Ebola viral disease vaccine (EVD) \citep{WatsonJones2022}. Some participants are recruited in the usual TND fashion as they present for care suffering from symptoms consistent with an Ebola diagnosis (and are thus tested); in addition, however, any test-positive identified in this fashion leads to immediate testing for Ebola for all close contacts of the test-positive who are likely asymptomatic at that point. We examine a simple approach to estimate the common efficacy of the vaccine intervention based on these two sources of test positives and test negatives, complemented by an assessment of whether efficacy is the same for both sources. The EVD trial was not completed for the fortunate reason that the prevailing disease outbreak ended; nevertheless, the approach here will be important if this trial is ever recommenced or similar trials are conducted in the future}



\maketitle

\renewcommand\thefootnote{}

\renewcommand\thefootnote{\fnsymbol{footnote}}
\setcounter{footnote}{1}
\section{Introduction}

The test-negative design (TND) has become common in studies of infectious disease interventions, and many TND studies have informed policy in the past decade. For example, policies informed by TND studies have included annual seasonal influenza vaccine recommendations \citep{grohskopf}, implementation of a three-dose pneumococcal conjugate vaccine (PCV) schedule in the UK and other countries \citep{whitney}, recommendations to remove upper age restrictions for receipt of rotavirus vaccines among children, and administration of oral cholera vaccines as a single dose during campaigns in emergency settings \citep{icg,hsiao,qadri}.  

The TND can be viewed as a version of a case-cohort study in which symptomatic
patients meeting predefined inclusion/exclusion criteria
are enrolled and subsequently classified as test-positive “cases”
or test-negative “controls” based on the results of definitive
diagnostic testing for the outcome under study. At the same time, eligible participants are classified as exposed or unexposed to an intervention of interest such as vaccination. As noted, this design is frequently used for evaluating
the effectiveness of seasonal influenza vaccination \citep{jackson2013test, foppa2013, sullivan2016theoretical, grohskopf}, and
its internal validity has been explored in depth \citep{orenstein2007, jackson2013test, deserres2013, sullivan2014, haber2015, sullivan2016theoretical, westreich2016beware}.
Briefly, validity depends primarily upon the avoidance of selection
bias in the sampling of cases and controls, as well as the
the extent to which the exposure distribution among controls is
representative of the exposure distribution among the source
population that gives rise to cases. Key assumptions and their
relevance to test-negative studies are discussed briefly below. For ease of
discussion we refer to the outcome of those testing positive for
the pathogen of interest as “test-positive illness” and in those testing
negative as “test-negative illness.”

The key assumptions for unbiased estimation of an intervention effect using data from a standard TND study include that (i) test-negative illness occurrence is not associated with the intervention, (ii) the relative propensity of individuals to seek health care is non-differential by outcome status, (iii) the test used to determine disease status is highly sensitive and specific, i.e. no misclassification, (iv) the sampling of test-negatives is unfiltered; that is, controls must be sampled from the whole population at risk without excluding those who test-positive at another time during the study period, (v) the efficacy of the intervention is not modified by healthcare-seeking behavior, and (vi) participants with test-negative illness are recruited only when test-positive illness is circulating. The first of these assumptions requires a careful definition of what is considered to be a suitable test-negative illness that also must exhibit non-specific symptoms similar to those that occur with the test-positive illness. Assumption (ii) is plausible if tested individuals are not aware of their (likely) outcome status prior to seeking care and testing. Assumption (iii) can be weakened by appropriate statistical adjustments. Assumptions (iv) and (vi)   
guarantee that test-negative controls are from the source population \textit{at risk} for the test-positive illness. Finally, assumption (v) allows extrapolation from the healthcare-seeking population to the general population.

Because cases (test-positives) and controls (test-negatives) are recruited from the same patient population, and restricted to those
seeking care, the design was intended to ameliorate if not eliminate, bias caused by health-care-seeking
behavior \citep{jackson2013test, haber2015, sullivan2016theoretical}. 

We continue our discussion with a definition of the causal estimand of interest for vaccine effectiveness studies, namely the (causal) population Relative Risk ($RR$) defined as the ratio of the incidence proportions of the disease of interest comparing the population under the two counterfactual conditions, the first where all members of the population are vaccinated and the other where no one is vaccinated. It is important to note here the specific definition of incidence being used. For the target estimand for a TND, we define incidence as both testing positive for the pathogen of interest and developing symptoms requiring health care. For convenience, we shall denote this true comparative population $RR$ by $\lambda$, again with the understanding that this efficacy is measured by a reduction in the likelihood of developing the specified symptoms necessary for inclusion. In Section 2, we also consider including asymptomatic infections (i.e. test-positives). 

Several authors have explored the statistical rationale and underlying assumptions of the traditional TND, showing that
the empirical odds ratio ($\widehat{OR}$)—i.e., the observed sample odds of exposure in test-positives relative to that for test-negative controls—targets $\lambda$ the population Relative Risk ($RR$), {\em in the healthcare-seeking population}, providing
underlying assumptions (noted above) are met \citep{jackson2013test, haber2015}. This is a direct consequence of the case-cohort nature of the sampling of test-positive and test-negative participants. We emphasize again that this definition of Relative Risk is specific to the definition of a test-positive outcome in the sense that it depends on the level of symptoms that are required to be eligible for recruitment. As such, this Relative Risk may vary if eligibility requires more serious symptoms such as hospitalization, etc. 

This causal estimand of interest as defined by potential outcomes ($RR$) can be equivalently assessed through the use of directed acyclic graphs as long as the standard causal assumptions of (i) consistency, (ii) no interference, and (iii) no unmeasured confounding are satisfied. The last of the these three assumptions requires that all relevant confounding variables are included in the relevant causal diagram.   The
causal structure underlying this approach has been studied in detail  \citep{sullivan2016theoretical}, with a particular focus on the fact that health care seeking behavior may not be a binary characteristic. For the traditional TND examining influenza,
exposure refers to seasonal vaccination so that an estimate of the RR for the outcome associated with exposure directly yields an estimate of vaccine effectiveness, $1-RR$. 

Throughout the statistical development described above, it has been assumed that both test positives and test negatives are ascertained through healthcare facilities where potential participants seek care because of undifferentiated symptoms so that such individuals are blinded to their disease status at the time of seeking care when their exposure status (i.e. vaccination) is also determined. However, in recent applications, this simple ascertainment process is not satisfied in that there may be multiple reasons why individuals seek testing (that ultimately determines whether they are classified as test-positives or test-negatives). For example, in recent Covid-19 test-negative studies \citep{andrejko2022}, individuals may seek, or be referred for, asymptomatic testing for screening purposes (for travel, etc), or because these individuals may have been exposed to the virus (by a family member, for example). 

\cite{lewnard2021theoretical}  pointed out that reasons for testing other than symptoms may introduce bias, and \cite{shi2023current} conceptualized this issue through a causal graph. Both papers, and \cite{vandenbroucke2022evolving}, state that a stratified analysis based on testing reasons could alleviate the bias, but no quantification or formal solution was given.

Here we focus on a specific example of an extended test-negative design with two distinct sources of participants, motivated by a proposed study to assess a vaccine for Ebola virus disease (EVD) in the eastern Democratic Republic of Congo \citep{WatsonJones2022, WHO2019, SAGE2019, J&J2019}. In this trial, a traditional test-negative design was employed but, in addition to traditional symptomatic testing, close contacts of any observed test-positives in the first stage were tested for the presence of the Ebola virus, prior to the onset of any symptoms. Vaccination status was ascertained for both symptomatic and asymptomatic participants at the time of recruitment.

\cite{pearson2022potential}  quantify the bias from a naive TND analysis based on data from this kind of hybrid design with two distinct reasons for testing: (i) symptomatic self-reporting or (ii) testing close contacts of known cases, and proposed a weighted average estimator to remove bias from naive aggregation. This required a strong assumption that the VE for either recruitment population is the same, an assumption we examine further below. 

A naive solution to filter out asymptomatic subjects and only use the remaining data to perform classical TND analyss was also discussed. Although this procedure is valid, it loses a large amount of information, yielding less efficient estimation.  In this paper, we discuss the analysis of data from this hybrid design and examine a simple method for assessing the shared efficacy of vaccine intervention in response to these two sources (symptomatic samples and case contact samples) of test positives and test negatives, together with an assessment of whether the efficacy is the same for both sources. The merits of our method are shown by comparing results with existing benchmark estimators.

\section{Traditional Test-Negative Data Supplemented by Testing of Close Contacts of Cases}

Before discussing the causal structures underlying the recruitment of both symptomatic and asymptomatic participants, we discuss briefly the role of an intervention such as vaccination in both preventing infections and specific symptoms. Following \cite{lewnard2021theoretical}, let $\lambda_S$ be the relative risk of {\em infection} for a vaccinated versus unvaccinated individual, resulting from a reduced susceptibility to the acquisition of a pathogen or accelerated clearance of the pathogen. Again, this causal parameter can be defined in terms of potential outcomes. Infection here is measured through a highly accurate diagnostic test even if no symptoms ever present. Further, define $\lambda_P$ as the relative risk of the infection meeting a particular severity threshold, e.g., clinical symptoms requiring care, hospitalization, intensive care unit admission, or mechanical ventilation) for an infected vaccinated individual versus an infected unvaccinated individual (where $P$ stands for progression), again using potential outcomes to stress the causl nature of the estimand. 

Then the overall relative risk (for the pre-defined symptom severity outcome) associated with vaccination is just $\lambda=\lambda_S \times \lambda_P$, with $VE =1-\lambda$.  As we discuss below, the traditional TND (with symptomatic recruits) estimates $\lambda$ but cannot separate out the distinct vaccine effects $\lambda_S$ and $\lambda_P$. Further, this Relative Risk is estimated only in the healthcare-seeking population from which participants are drawn absent assumption (v) discussed in the previous section--for clarity, we refer to this Relative Risk as $\lambda_H$. However,  as we elucidate below, testing of case contacts provides direct estimates of $\lambda_S$---which may differ from $\lambda$---in the entire population, whether healthcare-seekers or not. In particular, $\lambda = \lambda_S$ only if $\lambda_P = 1$, that is, the intervention---e.g. vaccination---has no influence on symptoms after infection has occurred. In short, for $\lambda_H = \lambda_S$, we require---at least---assumption (v) above and that $\lambda_P = 1$. We discuss additional potential confounding below.

We now elucidate this reasoning by exploiting two Directed Acyclic Graphs (DAGs) in Figure \ref{fig:consist_p} associated with test-negative designs for the two distinct reasons for testing.

\subsection{Test-Negative Design with Symptomatic Participants}\label{sec:sr} 

\begin{figure}[!p]
	\centering
	\begin{tabular}{cc}
		& \includegraphics[scale=0.45]{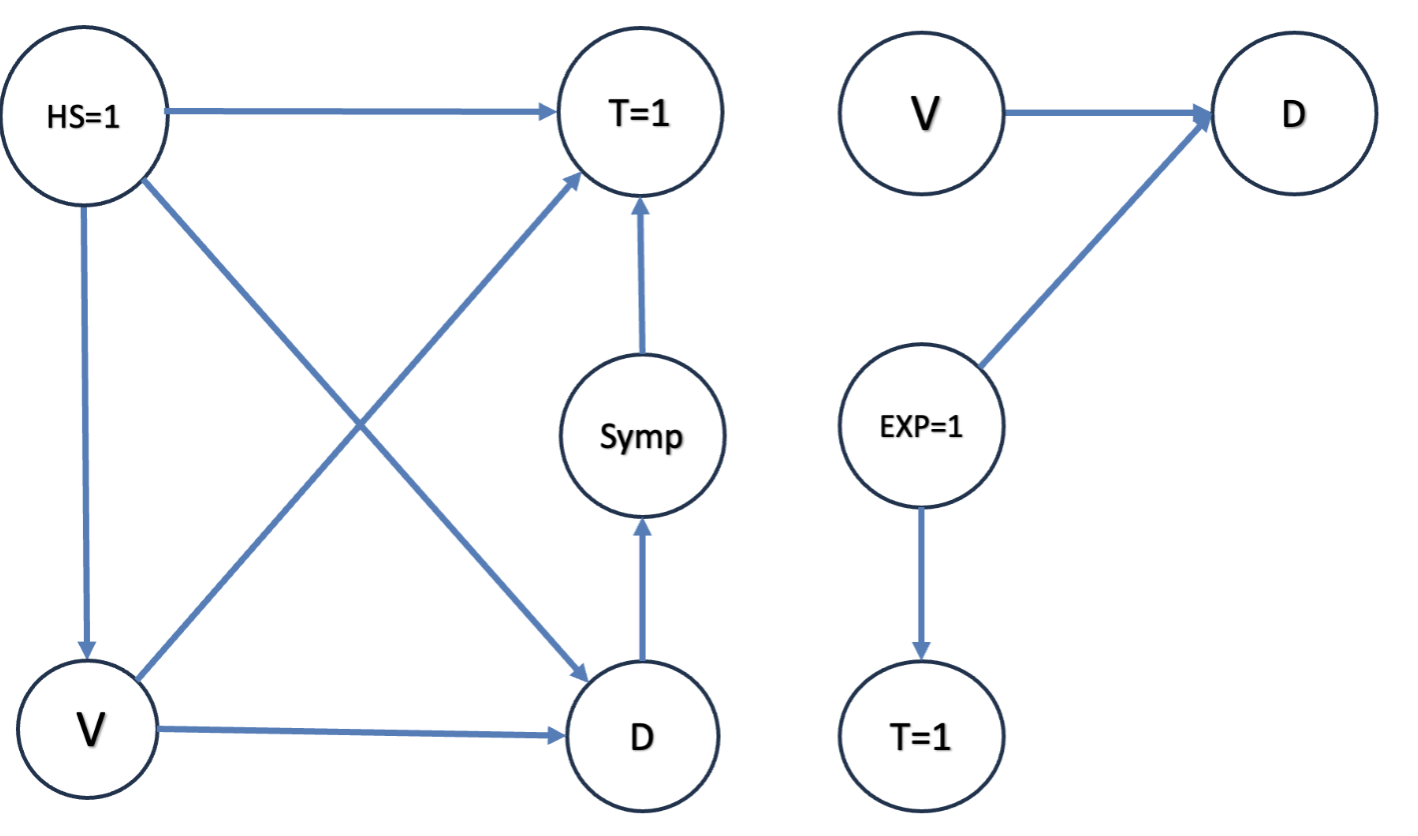}\\
		&(a)\qquad \qquad \qquad \qquad \qquad \qquad \qquad \quad(b)\\
	\end{tabular}
	\caption{DAGs for Self-Reported (Symptomatic) and Case Contact Tracing (Asymptomatic)}
	\label{fig:consist_p}
\end{figure}

The illustration on the left panel of Figure \ref{fig:consist_p} describes a DAG for a traditional test-negative design that recruits symptomatic patients only. These are individuals who proactively seek medical attention when they experience symptoms pertinent to the disease of concern (and the test-negative conditions for that matter). Within this population, the results of an appropriate diagnostic test are classified into two groups: 'test-positive' and 'test-negative' (control).

We define the several nodes, or variables, within this DAG, as follows:

\begin{itemize}
	\item $HS$: This node denotes the variable associated with quantifying healthcare-seeking behavior. For simplicity, we treat this variable as binary. Specifically, $HS=1$ indicates that the participant seeks care upon onset of the symptoms defined by the study protocol. Only such individuals can be recruited in a traditional test-negative study.
	\item V: This node refers to the vaccination, or intervention, status of the patient.
	\item D: This node signifies the disease status, elucidating whether the target infection has occurred or not (i.e., a test-positive result).
	\item Symp: This node is an indicator variable for the presence of disease-related (test-positive or test-negative) symptoms.
	\item $T=1$: This depicts whether a test has been conducted, essentially indicating that the participant is eligible and has been recruited for the study.
\end{itemize}

The arrow $V\rightarrow D$ is indicative of the fact that a patient's vaccination status can influence the likelihood of contracting the disease in question. There is no direct effect from $D\rightarrow T$ since taking the test is entirely determined by health-seeking behavior and symptoms. 

Next, we introduce an additional assumption regarding the relationship between vaccine effectiveness, healthcare-seeking behavior, and other (test-negative) illnesses that permit identification.

\begin{assumption}\label{self_rep}
	The effectiveness of the vaccine is not modified by healthcare-seeking behavior, and the vaccine status does not cause the presence (or not) of test-negative illnesses.
\end{assumption}

The second part of this assumption (discussed as Assumption (i) in Section 1) permits the identification of vaccine effectiveness, whereas the first part allows extrapolation of estimated vaccine effectiveness to the entire symptomatic population whether they seek care or not. These assumptions are also made in many previous test-negative design papers \citep{jackson2013test,pearson2022potential}. Note that, as discussed above, the traditional test-negative design cannot assess vaccine effectiveness in preventing infection, only on preventing {\em symptomatic} infection. We return to this later when we expand the design to include testing asymptomatic participants.

Identification of vaccine effectiveness  within the symptomatic tested population then follows:
\begin{align*}
	\textrm{OR}&=\frac{P(D=1|V=1,Symp=1,T=1,HS=1)}{P(D=0|V=1,Symp=1,T=1,HS=1)}\times \frac{P(D=0|V=0,Symp=1,T=1,HS=1)}{P(D=1|V=0,Symp=1,T=1,HS=1)}\\
	&=\frac{P(D=1,V=1,Symp=1,T=1,HS=1)}{P(D=0,V=1,Symp=1,T=1,HS=1)}\times \frac{P(D=0,V=0,Symp=1,T=1,HS=1)}{P(D=1,V=0,Symp=1,T=1,HS=1)}\\
	&=^{(a)}\frac{P(D=1,V=1,Symp=1,HS=1)}{P(D=0,V=1,Symp=1,HS=1)}\times \frac{P(D=0,V=0,Symp=1,HS=1)}{P(D=1,V=0,Symp=1,HS=1)}\\
	&=\frac{P(Symp=1|D=1,V=1,HS=1)\times P(D=1,V=1,HS=1)}{P(Symp=1|D=1,V=0,HS=1)\times P(D=1,V=0,HS=1)}\\&\qquad\times \frac{P(Symp=1,D=0,V=0,HS=1)}{P(Symp=1,D=0,V=1,HS=1)}\\
	&=^{(b)}\frac{ P(D=1,V=1,HS=1)}{ P(D=1,V=0,HS=1)}\times \frac{P(Symp=1,D=0,V=0,HS=1)}{P(Symp=1,D=0,V=1,HS=1)}\\
	&=\frac{P(V=1|D=1,HS=1)}{P(V=0|D=1,HS=1)}\cdot \frac{P(V=0|Symp=1,D=0,HS=1)}{P(V=1|Symp=1,D=0,HS=1)}\\
	&=^{(c)}\frac{P(V=1|D=1,HS=1)}{P(V=0|D=1,HS=1)}\times \frac{P(V=0|HS=1)}{P(V=1|HS=1)}\\
	&=\frac{P(D=1|V=1,HS=1)}{P(D=1|V=0,HS=1)}=\lambda_H =^{(d)}\lambda .
\end{align*}
There are several steps listed above that merit further explanation.
Specifically, equality $(a)$ holds since
\begin{align*}
	\frac{P(T=1|D=1,V=1,Symp=1,HS=1)}{P(T=1|D=0,V=1,Symp=1,HS=1)}= 1=\frac{P(T=1|D=0,V=0,Symp=1,HS=1)}{P(T=1|D=1,V=0,Symp=1,HS=1)}.
\end{align*}
Essentially, with a given vaccine status $V$, $Symp=1$, and healthcare-seeking behavior $HS=1$, the decision to test $T$ remains independent of the disease status $D$. This is a valid observation in the context of the symptomatic healthcare-seeking population, where the decision to pursue testing is primarily influenced by healthcare-seeking behavior, symptoms, and vaccine status. Disease status only influences testing behavior indirectly, through the manifestation of relevant symptoms. Note that in well-designed test-negative studies, this assumption is reasonable since participants do not know their exact disease status before testing. The assumption might be violated if, for example, other circumstances suggest disease status to the participant before testing (such as knowledge of close contact with the relevant pathogen of interest).

Equality $(b)$ holds since
\begin{align*}
	\frac{P(Symp=1|D=1,V=1,HS=1)}{P(Symp=1|D=1,V=0,HS=1)}=1.
\end{align*}
The underlying logic of this equation is that given a certain health-seeking behavior $HS$ and disease outcome $D$, the status of symptoms is directly dictated by the presence of disease ($D = 1$), and remains unaffected by vaccine status. This directly assumes that the vaccine does not prevent (or ameliorate) symptoms, or in symbols that $\lambda_P =1$;if infected by the test-positive pathogen, it is possible that prior vaccination may reduce or even eliminate symptoms (as has been postulated for Covid-19 vaccines). This is unlikely to be true for vaccines for Ebola viral disease, however.   

Equality $(c)$ is valid since per Assumption \ref{self_rep}, amongst control participants (i.e. test-negatives), the vaccine status is independent of symptoms, and its relative frequency reflects that of the general healthcare-seeking population. 

Finally, Assumption \ref{self_rep} posits that the result of the intervention is not modified by the health-seeking behavior, validating Equality $(d)$. Consequently, we derive the following expression:
\begin{align*}
	\frac{P(D=1|V=1,HS=1)}{P(D=1|V=0, HS=1)}=\frac{P(D=1|V=1)}{P(D=1|V=0)}=\lambda.
\end{align*}

Based on the identification assumptions, we are thus able to leverage the symptomatic participant data on vaccination to ascertain the true effectiveness of the vaccine. We now briefly outline the sampling context for the traditional test-negative design used for recruiting individuals. Table \ref{tab-sr} provides the breakdown of both test-positives ($D$) and test-negatives ($D^c$) by their vaccination status. In this table, $A'$ denotes the observed number of vaccinated individuals in the test-positive cases in the test-negative design sample, etc. This yields the following (conditional) likelihood, where we use the subscript $SR$ to denote the data arising from symptomatic recruited patients:

\begin{equation}\label{eq_selfreport}
	\mathcal{L_{SR}} \propto p_{0}^{B'}(1-p_{0})^{D'}p_1^{A'}(1-p_1)^{C'}     
\end{equation}
where $p_0=p(V)$ and $p_1=P(V|D)$ with $ \frac{p_1}{1-p_1} = \frac{p_0}{1-p_0}\cdot \lambda$. Note that, from above, we do not need to condition on $HS$ by Assumption \ref{self_rep}; also, $p_0$ does not condition on $D^c$ due to the case-cohort nature of the sampling; that is, controls may also be sampled as cases.
In this context, $A',B', C'$, and $ D'$ represent the number of observed test samples in distinct categories from symptomatic participants.

\subsection{Testing of Contacts of Positives (Cases)}\label{sec:ct}
The right-hand panel of Figure 1 illustrates a Directed Acyclic Graph (DAG) for data that arises from testing close contacts of test-positives. In this population, the testing behavior, as per the DAG, is solely determined by the exposure status ($EXP$), where $EXP = 1$ indicates that an individual is a close contact with a known person who has tested positive for the presence of $D$; the decision to test is unaffected by the individual's vaccine status $(V)$ and disease status $(D)$, given $EXP=1$. Such a DAG might describe scenarios involving severe diseases, such as Ebola, where individuals who have had close contact with confirmed positive cases are expected to undergo immediate (asymptomatic) testing. Consequently, we make the following assumption:

\begin{assumption}\label{assump_cc1}
	Any individual  identified as a close contact is obligated to undertake a screening test for the presence of $D$, that is,
	\begin{align*}
		P(T=1|EXP=1)=1.
	\end{align*}
\end{assumption}

This assumption follows the protocol of the Ebola vaccine trial \citep{pearson2022potential}. 

Next, we propose an additional assumption that the efficacy of the vaccine is independent of the exposed population.

\begin{assumption}\label{assump_cc2}
	The effectiveness of vaccine is identical among the exposed and unexposed (on the risk ratio scale).
\end{assumption}

This assumption is not about potential confounding but homogeneity of effect, stating that the causal relative risk amongst contacts of known cases does not differ from the analogous causal relative risk in individuals outside of the study. Note that such individuals who become infected must have experienced a case contact at some point. In addition, in most vaccine studies, there is often little evidence of vaccine effectiveness varying by population subgroups (for Covid-19, the original trials of the Pfizer and Moderna vaccines considered subgroup analyses). A major exception to this general statement concerns immunocompromised patients where vaccines often provide weaker protection, a situation where there is a strong biological rationale to expect a difference. This comment also supports the first part of assumption 1 in Section 2.1 regarding homogeneity of vaccine effectiveness across differing healthcare-seeking subgroups.

Under these two assumptions, and using $RR$ to denote the relative frequency of positive test results among the tested contacts,  we have
\begin{align*}
	\textrm{RR}&\equiv \frac{P(D=1|V=1,EXP=1,T=1)}{P(D=1|V=0,EXP=1,T=1)}\\
	&=^{(e)}\frac{P(D=1|V=1,EXP=1)}{P(D=1|V=0,EXP=1)}\\
	&=^{(f)}\frac{P(D=1|V=1)}{P(D=1|V=0)}=\lambda_S
\end{align*}

Equality $(e)$ follows from Assumption \ref{assump_cc1}, which posits that all close contacts undergo testing given exposure to confirmed positive cases. Furthermore, Equality $(f)$ is valid due to Assumption \ref{assump_cc2}, which assumes that the efficacy of the vaccine is independent of exposure status. Note that, in this context, the variable $D=1$ requires only a positive test for infection with the pathogen and does not require the presence of any symptoms. Recall that $\lambda_S = \lambda$ if $\lambda_P =1$.

Given this identifiability, we consider the following (conditional) likelihood arising from Table \ref{tab-ct} data that classifies contacts' test results by outcome and vaccination status of the contact. In this table, $A^*$ counts the number of exposed contacts who both test positive at recruitment and are vaccinated, etc.
\begin{equation}\label{eq_contacttrace}
	\mathcal{L_{CT}} \propto q_{0}^{C^*}(1-q_{0})^{D^*}q_1^{A^*}(1-q_1)^{B^*}     
\end{equation}
where $q_0=p(D|V^c)$ and $q_1=P(D|V)= \lambda_S q_0.$

There is a tacit assumption made here that any subsequent intervention on contacts of known cases does not affect their chances of infection as measured at the time of recruitment (although it might improve symptoms after any infection). Further, we assume temporality in that contacts who test-positive are assumed to have followed infection of the case and not the reverse, this assumption being based on the earlier appearance of symptoms in the case.

\subsection{Combined Likelihood}
From the discussion of Sections \ref{sec:sr} and \ref{sec:ct}, we note the presence of two distinct likelihood functions, each capable of evaluating the effectiveness of the vaccine using observed samples, albeit estimating somewhat different measures of efficacy absent further assumptions. If we invoke the assumptions discussed above, including that $\lambda_P = 1$, and we assume that the two estimands are equivalent for this and other reasons (we turn to a discussion of confounding later in this section), we propose a straightforward methodology to assess the collective efficacy of the vaccine intervention by aggregating the two distinct sources of information. If we use a composite likelihood, the full (conditional) likelihood is then simply $\mathcal{L}=\mathcal{L_{SR}} \times \mathcal{L_{CT}}$.  When contacts of cases from the initial TND are tested (as was planned for the motivating Ebola trial), the use of this combined likelihood makes an additional assumption, namely that the outcome of a case contact is independent of the characteristics of the associated case. This assumption can be weakened if covariates are introduced into the analysis (as discussed further below). 

The negative logarithm of this combined likelihood, $\mathcal{L}=\mathcal{L}_{SR}\times \mathcal{L}_{CT}$ simplifies to
\begin{align}
	\log \mathcal{L}&=-\Big[B'\log p_0+D'\log (1-p_0)+A'\log (p_1)+C'\log(1-p_1) \nonumber \\&\quad +C^*\log q_0+D^*\log(1-q_0)+A^*\log (q_1)+B^*\log (1-q_1)\Big],\label{combined_likelihood}
\end{align}
based on the data from Tables \ref{tab-sr} and \ref{tab-ct}. Furthermore, from sections \ref{sec:sr} and \ref{sec:ct}, we substitute $p_1$ with $p_0\lambda/(1-p_0+p_0\lambda)$ and replace $q_1$ with $\lambda q_0$. Our primary interest lies in conducting statistical inferences for the common $\lambda$.

We note that the parameters $p_0, p_1, q_0, q_1$ are constrained within the interval $[0,1]$, whereas the parameter $\lambda$ is confined to the non-negative real numbers. These restrictions pose challenges for optimization procedures due to the necessity of operating within a constrained parameter space. To address this, we use a simple reparameterization. Specifically, we redefine $p_0$ as $1/(1+\exp(p_0'))$, $q_0$ as $1/(1+\exp(q_0'))$, and $\lambda$ as $\exp(\lambda')$. 


Then, employing this reparameterization in the combined likelihood \eqref{combined_likelihood} yields
\begin{align*}
	\mathcal{L}'&=-\log \mathcal{L}(p_0',q_0',\lambda')\\&=-\bigg[B'\log \bigg(\frac{1}{1+\exp(p_0')}\bigg)+D'\log \bigg(\frac{\exp(p_0')}{1+\exp(p_0')}\bigg)+A'\log \bigg(\exp(\lambda')\cdot \frac{1}{\exp(p_0')+1}\bigg)\\&\qquad-A'\log \bigg(\exp(\lambda')\cdot \frac{1}{\exp(p_0')+1}+\frac{\exp(p_0')}{1+\exp(p_0')}\bigg)+C'\log \bigg(\frac{\exp(p_0')}{1+\exp(p_0')}\bigg)\\&\qquad-C'\log\bigg( \exp(\lambda')\cdot \frac{1}{\exp(p_0')+1}+\frac{\exp(p_0')}{1+\exp(p_0')}\bigg)+C^*\log\bigg(\frac{1}{\exp(q_0')+1}\bigg)\\&\qquad+D^*\log\bigg(\frac{\exp(q_0')}{1+\exp(q_0'))}\bigg)+A^*\log\bigg(\exp(\lambda')\frac{1}{1+\exp(q_0')}\bigg)\\&\qquad+B^*\log\bigg(1-\exp(\lambda')\cdot\frac{1}{1+\exp(q_0')}\bigg)\bigg].
\end{align*}

Subsequently, assuming that the true value of $(p_0, q_0,\lambda )$ does not lie on the boundary of the parameter space, it is then straightforward to establish the existence of a unique maximum likelihood estimate, denoted by $\hat \lambda_c$.

We end this section by noting that, of course, the two likelihood functions \eqref{eq_selfreport} and \eqref{eq_contacttrace} can be maximized separately to provide estimates of $\lambda$ and $\lambda_S$. With this approach, a likelihood ratio test can be used to test the equality of the two underlying Relative Risks. Note that rejection of the null hypothesis, in this case, does not formally establish that $\lambda_P =1$ since there are other potential explanations of why the two Relative Risks may differ. For example, the exposure experiences of close contacts of cases may differ from those recruited symptomatically from the general population which may affect the performance of a vaccine intervention. Further, there is the issue of confounding by healthcare-seeking behavior amongst case contacts. Nevertheless, this likelihood ratio test may provide some evidence that a vaccine influences the progression of symptoms after breakthrough infections.

To address potential confounding, it is important to expand the likelihood functions \eqref{eq_selfreport} and \eqref{eq_contacttrace} to incorporate the influence of relevant covariates. The choice of covariates might differ for the distinct likelihoods as varying covariate information might be available from the two data sources. Particularly important here is the role of healthcare-seeking behavior. This variable is naturally controlled in the estimation of $\lambda$ from the first likelihood (since sampling is restricted to healthcare-seekers by design), $\mathcal{L_{SR}}$, but is not addressed by naive use of $\mathcal{L_{SR}}$. In this case, measurement and adjustment for covariates that act as proxies for healthcare-seeking behavior may be especially important. When covariates are included (through the use of appropriate generalized linear models (GLMs) in both likelihood components), the estimand reflects a conditional Relative Risk and thus vaccine effectiveness. Given the collapsibility of the Relative Risk, this will still yield a marginal Relative Risk associated with $V$ so long as there is no interaction between $V$ and any of the included covariates. Note that---for most vaccines---there is usually no biological basis for expecting such interactions to occur (other than perhaps for measures of immune compromise).  

Of further interest is the vaccination status of the case associated with contacts in data underlying likelihood function \eqref{eq_contacttrace}. Information on the vaccination status of the assumed infected person (the individual who causes the infection, or not, in a contact) should be available for contacts of known cases (particularly if the cases arise from symptomatic recruitments in the underlying TND study). It is unlikely, however, that this is known for the original TND participants since, in most cases, the source of their infection is likely unknown. Analysis of this covariate in an expanded version of \eqref{eq_contacttrace} would allow an assessment of whether vaccination reduces infectiousness in a vaccinated individual who experiences a breakthrough infection. 

Note that adjustment for covariates may also control for any source of dependence between a case (used in likelihood \eqref{eq_selfreport}) and contacts (used in likelihood \eqref{eq_contacttrace}) which has been ignored in use of the combined likelihood as noted above.

\section{Numerical Studies}




In this section, we assess the performance of the combined maximum likelihood estimator through simulations. For simplicity, we consider sample sizes of $n=500$ with $(B'+D')=(A'+C')=(C^*+D^*)=(A^*+B^*)=n$ \footnote{A similar conclusion can be drawn when employing varying sample sizes for different data sources.}.

We selected the true parameters $p_0^*=0.8$ and $q_0^*=0.6$ while varying the value of $\lambda$ across the range $0.6\sim 1.0$, reflecting high rates of vaccination and moderate values of efficacy.  We provide a detailed implementation of the optimization procedure in the appendix. 


We compare the performance of the estimator $\hat\lambda_c$, obtained by minimizing the combined loss function in \eqref{combined_likelihood}, with additional benchmark estimators. In simulating the data, we allow for two situations: (i) the vaccine effects of the two sources are the same, and (ii) the vaccine effects from the two sources are different, for example, whether $\lambda = \lambda_S$ or not. 

For each case, we study four benchmarks involving different weighted averages of the two estimators based on separate maximizations of \eqref{eq_selfreport} and \eqref{eq_contacttrace}, denoted as $(w_1\hat\lambda_1+w_2\hat\lambda_2)$, where $\hat\lambda_1=(\frac{\hat p_1}{1-\hat p_1})/(\frac{\hat p_0}{1-\hat p_0})$ and $\hat\lambda_2=\hat q_1/\hat q_0$. Here, $\hat p_0, \hat p_1, \hat q_0, \hat q_1$ represent the separate maximum likelihood estimators based on \eqref{eq_selfreport} and \eqref{eq_contacttrace}, respectively. It is noteworthy that a weighted estimator was previously suggested by \cite{pearson2022potential}; however, the specific choice of weights was not explicitly discussed. Therefore, we compare estimation results using $\hat\lambda_c$ with (i) the traditional TND estimator where we set $w_1=1$ (benchmark 1) and (ii) a simple average (with $w_1=1/2,w_2=1/2$) (benchmark 2) and (iii) a third benchmark (benchmark 3) where the weights are proportional to the (estimated) inverse standard errors of $\hat\lambda_1$ and $\hat\lambda_2$. In addition, we also compare $\hat\lambda_c$ with an odds ratios estimator based on a naive aggregation of the two distinct testing samples from the two sources (the fourth benchmark); that is, the estimator $[(A'+A^*)\times(D'+D^*)]/[(B'+B^*) \times (C'+C^*)].$


For each assumed value of $\lambda$, we generate 500 simulated data sets, reporting the average value of our estimator $\hat\lambda_c$ as well as the standard error of $\hat\lambda_c$. Detailed simulation results are presented in Table \ref{tab1}. We also consider scenarios where the vaccine effects from the two testing sources are different (e.g. when $\lambda \neq \lambda_S$). In this case, we study the performance of the aforementioned estimators and also conduct a likelihood ratio test with significance level $\alpha=0.05$ on the equivalence of the vaccine effect from two sources. The results are presented in Tables \ref{tab2} and  \ref{tab_twosample}, respectively. 


\subsection{Results}\label{sec:simu1}

Based on Table \ref{tab1}, the simple aggregated odds ratio (without differentiating the testing sources/reasons; that is, Benchmark 4) exhibits bias. This occurs even though the vaccine effects from the two distinct testing sources are identical, leading to a failure to accurately identify the true vaccine effect. In contrast, when data from the two testing sources are separately analyzed, our proposed estimator---derived from the combined likelihood of Equation \ref{combined_likelihood}---demonstrates greater efficiency as compared to the other benchmark estimators. This assertion is further corroborated by the theoretical findings presented in the appendix, wherein we establish that the variance of the proposed estimator achieves the Cramér–Rao lower bound. 

Table \ref{tab2} reveals that when the vaccine effects from the two testing sources differ ($\lambda \ne \lambda_S$), all combined estimators exhibit bias. Note that the simulations here reflect a possible scenario where $\lambda_1 < \lambda_2 =1$, so the vaccine here does not reduce the risk of infection but reduces the risk of symptoms post-infection.  Therefore, in reality, it is important to first assess the comparability of vaccine effects from the distinct data sources, for example by implementing a likelihood ratio test. To this end, we present the results of conducting such a likelihood test in Table \ref{tab_twosample}. When the null hypothesis holds, the type I error is well controlled under $5\%$, and when the alternative holds, the power rapidly increases to $1$ when the number of testing samples grows.

In principal, contacts of known cases for whom infection is observed could be followed to detect onset of symptoms. This would allow a direct assessment of whether the vaccination status of an infected individual affects symptom development. However, in practice, such an assessment is not likely to reflect the role of vaccination in general as other interventions that may influence symptoms will likely be available to infected contacts immediately upon determination of their infection.

\section{Discussion}
Data from the motivating  Ebola vaccine trial is not available for analysis for the very best of reasons, namely that the 2018 Ebola outbreak in the Democratic Republic of the Congo ended. Nevertheless, the methodology developed here will be valuable if the vaccine trial is ever reopened, or if similar vaccine trials with this design are planned in the future. We have shown here that there is a gain in efficiency in using the hybrid design although, as with any observational study, and the traditional test-negative design in particular, this comes with the requirement of additional assumptions. The hybrid design also permits careful consideration of whether a vaccine only ameliorates symptoms or whther it also protects against infection.

Our methodological assessment of the motivating Ebola vaccine study example with two sources of testing data illustrates the importance of considering the reasons for testing and avoiding any temptation to aggregate testing data naively. Further, it is crucial to understand that data arising from different testing sources may estimate different measures of effectiveness. The design of such hybrid TND studies must carefully consider an appropriate discussion of symptoms for eligibility as well as the possibility that the intervention may alleviate the development of such symptoms separately from any effect on infection. In most cases, it will also be important to pre-define covariates for measurement that will allow adjustment for potential confounding by healthcare-seeking behavior when implementing asymptomatic testing. When the estimands associated with different testing sources differ, further information will usually be required to disentangle the root causes of such differences.

\begin{table}[!p]
	\centering
	\def\arraystretch{1.8}
	\setlength\tabcolsep{16pt}
	\begin{tabular}{c c c}
		\toprule
		&$V$ & $V^c$   \\
		\hline 
		$D$ &$A'$ &$C'$	
		\\
		\hline
		$D^c$	&$B'$ &$D'$\\
		\bottomrule\\
	\end{tabular}
	\caption{Test samples from the self-reported population. In this context, the rows $D$ and $D^c$ represent test-positive and test-negative groups, respectively. The columns $V$ and $V^c$ denote whether individuals within a specific group (positive or negative) have been vaccinated.}
	\label{tab-sr}
\end{table}

\begin{table}[!p]
	\begin{center}
		\def\arraystretch{1.8}
		\setlength\tabcolsep{16pt}
		\begin{tabular}{c c c}
			\toprule
			&$V$ & $V^c$   \\
			\hline 
			$D$ &$A^*$ &$C^*$	
			\\
			\hline
			$D^c$	&$B^*$ &$D^*$\\
			\bottomrule\\
		\end{tabular}
	\end{center}
	\caption{Test sample counts from testing close contacts of positives. The definition of $D,D^c,V,V^c$ are consistent with those in Table \ref{tab-sr}}
	\label{tab-ct}
\end{table}

\begin{table}[h]
	\begin{center}
		\def\arraystretch{1.2}
		\setlength\tabcolsep{4pt}
		\begin{tabular}{cc ccccc}
			\toprule
			& & Proposed Estimator & Benchmark 1 & Benchmark 2 &Benchmark 3 &Benchmark 4 \\
			\hline 
			\multirow{2}{*}{$\lambda=1.0$	}			 &Averages & 1.001 & 1.019 & 0.879 &1.010 &1.005 \\ 
			&SDs  &0.049 & 0.163 & 0.085 & 0.055 & 0.069 \\
			
			\multirow{2}{*}{$\lambda=0.9$	}			 &Averages &0.902 & 0.904 &0.904 &0.903 &0.773 \\ 
			&SDs  &0.048 &0.136 &0.074 &0.053 &0.073\\
			\multirow{2}{*}{$\lambda=0.8$	}			 &Averages &0.799 &0.808 &0.804 &0.802 &0.673 \\ 
			&SDs &0.043 &0.122 &0.065 &0.047 &0.060 \\
			\multirow{2}{*}{$\lambda=0.7$	}		 &Averages &0.700 &0.704 &0.703 &0.702 &0.584 \\ 
			&SDs  &0.041 &0.108 &0.058 &0.045 &0.054 \\
			\multirow{2}{*}{$\lambda=0.6$	}	 &Averages &0.601 
			&0.609 &0.605 &0.603 &0.500 \\
			&SDs  &0.038 &0.092 &0.051 &0.041 &0.047 \\
			\bottomrule\\
		\end{tabular}
	\end{center}
	\caption{Results on the estimation of vaccine effectiveness $(1-\lambda)$ when the two testing sources share the same vaccine effect. Proposed Estimator: The estimator maximizes the combined likelihood. Benchmark 1: the traditional TND estimator based on symptomatic participants only. Benchmark 2: Simple average of $\hat\lambda_1$ and $\hat\lambda_2$. Benchmark 3: Weighted average of $\hat\lambda_1$ and $\hat\lambda_2$, with weights  proportional to the (estimated) inverse standard deviation of $\hat\lambda_1$ and $\hat\lambda_2$, respectively. Benchmark 4: The odds ratio estimator aggregating data from the two testing sources (symptomatic and case contacts). SDs provide the observed standard errors of the estimators over the 500 simulated samples
	}
	\label{tab1}
\end{table}

\begin{table}[h]
	\begin{center}
		\def\arraystretch{1.2}
		\setlength\tabcolsep{4pt}
		\begin{tabular}{cc ccccc}
			\toprule
			& & Proposed Estimator & Benchmark 1 & Benchmark 2 &Benchmark 3 &Benchmark 4 \\
			\hline 
			\multirow{2}{*}{$\lambda_1=0.9$	}			 &Averages & 0.979 & 0.910 & 0.956 & 0.977 & 0.849 \\
			&SDs & 0.049	& 0.145 &0.078 & 0.055 & 0.079 \\
			\multirow{2}{*}{$\lambda_1=0.8$	}			 &Averages & 0.963 & 0.801 & 0.900 & 0.941 & 0.811 \\
			&SDs & 0.047	& 0.122 & 0.067 & 0.051 & 0.075 \\
			\multirow{2}{*}{$\lambda_1=0.7$	}		 &Averages & 0.959 & 0.702 & 0.851 & 0.901 & 0.775 \\
			&SDs &0.045 & 0.101 & 0.056 &  0.047 & 0.068 \\
			
			\multirow{2}{*}{$\lambda_1=0.6$	}	 &Averages & 0.943 & 0.605 & 0.804 & 0.856 &  0.735\\
			&SDs &0.047	&0.090 &0.052 &0.048 & 0.069 \\
			\multirow{2}{*}{$\lambda_1=0.5$	}	 &Averages &0.917 & 0.504 & 0.751 & 0.795 & 0.679 \\
			&SDs & 0.046 & 0.074 & 0.047 & 0.046 & 0.063 \\
			\bottomrule\\
		\end{tabular}
	\end{center}
	\caption{Results on the estimation of vaccine effectiveness ($1-\lambda$) when the two testing sources do not share the same vaccine effect. The vaccine effect in the TND population (symptomatic recruitment) takes values $\lambda_1\equiv \lambda$ varying between $0.5\sim 0.9.$ In addition, we let $\lambda_2 \equiv \lambda_S=1.0$ be the Relative Risk amongst the case contact population. Proposed Estimator: The estimator maximizes the combined likelihood. Benchmark 1: the traditional TND estimator based on symptomatic patients only. Benchmark 2: Simple average of $\hat\lambda_1$ and $\hat\lambda_2$. Benchmark 3: Weighted average of $\hat\lambda_1$ and $\hat\lambda_2$, with weights proportional to the (estimated) inverse standard deviation of $\hat\lambda_1$ and $\hat\lambda_2$ respectively. Benchmark 4: The odds ratio estimator aggregating data from the two testing sources (symptomatic and case contacts). SDs provide the observed standard errors of the estimators over the 500 simulated samples}
	\label{tab2}
\end{table}

\begin{table}[!p]
	\begin{center}
		\def\arraystretch{1.4}
		\setlength\tabcolsep{14pt}
		\begin{tabular}{ccc}
			\toprule
			&   & Power or Type-I Error \\
			\hline 
			\multirow{4}{*}{$n=200$	}	
			& $\lambda_2=1.0, \lambda_1=1.0$		    &0.049  \\
			& $\lambda_2=1.0, \lambda_1=0.8$		    &0.164  \\
			&$\lambda_2=1.0, \lambda_1=0.6$	
			& 0.636  \\
			&$\lambda_2=1.0, \lambda_1=0.4$   & 0.968 \\
			\hline
			\multirow{4}{*}{$n=400$	}
			& $\lambda_2=1.0, \lambda_1=1.0$	    & 0.049 \\
			&  $\lambda_2=1.0, \lambda_1=0.8$	     & 0.268 \\
			&$\lambda_2=1.0, \lambda_1=0.6$  & 0.827 \\
			&$\lambda_2=1.0, \lambda_1=0.4$   & 1.000 \\
			\hline
			\multirow{4}{*}{$n=600$	}
			&  $\lambda_2=1.0, \lambda_1=1.0$	    & 0.045 \\& $\lambda_2=1.0, \lambda_1=0.8$			    & 0.341 \\
			&$\lambda_2=1.0, \lambda_1=0.6$	  & 0.943 \\
			&$\lambda_2=1.0, \lambda_1=0.4$  & 1.000 \\
			\bottomrule\\
		\end{tabular}
	\end{center}
	\caption{Results of Likelihood Ratio Test for $\lambda_1 \equiv \lambda =\lambda_2 \equiv \lambda_S$ at significance level $\alpha=0.05$. The Relative Risk of the case contact population is set at $\lambda_2 \equiv \lambda_S=1.0,$ with the Relative Risk of the TND population (symptomatic recruitment) $\lambda$ varying among $\{0.4,0.6,0.8,1.0\}$. Here, overall sample size is determined by $n= (B'+D')=(A'+C')=(C^*+D^*)=(A^*+B^*)$.   }
	\label{tab_twosample}
\end{table}

\section*{Acknowledgment}
This research was supported by grant \#R01-AI148127 from the National Institute for Allergy and Infectious Diseases. The authors would like to thank Dylan S. Small and Bingkai Wang for their insightful comments and discussion on this paper.

\section*{Supplementary Material}
The supplementary material provides the implementation on optimizing the combined likelihood and providing theoretical guarantees for the optimizer associated with likelihood \eqref{combined_likelihood}.

\appendix

\section*{Algorithm} \label{sec:alg}

Notably, if we independently solve the two distinct loss functions presented in sections 2.1 and 2.2, respectively, we arrive at a consistent estimator for the parameters $(p_0',q_0',\lambda')$. This consistent estimator can be selected as the initialization point for our gradient descent algorithm. This strategy leads to the first phase of the algorithm, which is concerned with initialization, as detailed in the subsequent discussion.:

Step 1, Initialization: We first solve $\hat p_0^{(0)},\hat p_1^{(0)},\hat q_0^{(0)},\hat q_1^{(0)}$ by maximizing likelihoods (2.1) and (2.2), respectively. We then initialize the re-parameterized parameters at the zeroth (initial) step $(\hat p_0'^{(0)},\hat q_0'^{(0)},\hat\lambda'^{(0)}).$  In specific, we first let $\hat p_0'^{(0)}=\log (1-1/\hat p_0^{(0)}),\hat q_0'^{(0)}=\log (1-1/\hat q_0^{(0)})$ and initialize the estimator of $\hat\lambda'^{(0)}$ by an log- weighted average of $\hat\lambda_1,\hat\lambda_2:$ $\log(w_1\hat\lambda_1+w_2\hat\lambda_2),$ where $\hat\lambda_1= (\frac{\hat p_1^{(0)}}{1-\hat p_1^{(0)}})/(\frac{\hat p_0^{(0)}}{1-\hat p_0^{(0)}})$ and $\hat \lambda_2=\hat q_1^{(0)}/\hat q_0^{(0)}$. The weights $w_1,w_2$ with $w_1+w_2=1$ are pre-determined. It can be chosen equally or proportional to the standard deviation of $\hat\lambda_1$ and $\hat\lambda_2,$ respectively.

Following the initialization phase, we progress to the subsequent stage, which entails the execution of the gradient descent algorithm. The specifics of this step are elucidated in the following description.

Step 2, Run Gradient Descent: Choose stepsize $\eta$ and run gradient descent on the parameters $p_0',q_0',\lambda'.$
The updates are presented as follows:
\begin{align*}
	(\hat p_0'^{(t+1)},\hat q_0'^{(t+1)},\hat\lambda'^{(t+1)})=(\hat p_0'^{(t)},\hat q_0'^{(t)},\hat\lambda'^{(t)})-\eta \Big(\nabla_{p_0'} \mathcal{L}(\theta^{(t)}),\nabla_{q_0'} \mathcal{L}(\theta^{(t)}),\nabla_{\lambda'} \mathcal{L}(\theta^{(t)})\Big)
\end{align*}
where $\theta^{(t)}=(\hat p_0'^{(t)},\hat q_0'^{(t)},\hat\lambda'^{(t)}).$
The gradient is given as follows:
\begin{align*}
	\nabla_{p_0'} \mathcal{L}(\theta)&=-\bigg[(C'+D')-(B'+D')\frac{\exp(p_0')}{1+\exp(p_0')}-(A'+C')\frac{\exp(p_0')}{\exp(\lambda')+\exp(p_0')}\bigg]\\
	\nabla_{q_0'} \mathcal{L}(\theta)&=-\bigg[(A^*+B^*+C^*+D^*)\frac{\exp(q_0')}{1+\exp(q_0')}-D^*-\frac{B^*\exp(q_0')}{1+\exp(q_0')-\exp(\lambda')}\bigg]\\
	\nabla_{\lambda'} \mathcal{L}(\theta)&=-\bigg[A'-\frac{(A'+C')\exp(\lambda')}{\exp(\lambda')+\exp(p_0')}+A^*-\frac{B^*\exp(\lambda')}{1+\exp(q_0')-\exp(\lambda')}\bigg],
\end{align*}
where $\theta=(p_0',q_0',\lambda').$

Step 3, Stopping Criterion: Upon reaching a point in the iterative process denoted as $\theta^{(t)}$, we terminate the algorithm when the infinity norm of the gradient at that particular iterate falls below a predetermined threshold value, denoted as $\tau$.

In the simulation experiments, we choose a step size of $\eta=0.01/n$, and a stopping threshold of $\tau=10^{-5}$ is used. 

\section*{Theoretical Guarantees}
In the previous simulation example, we observed that our estimator outperformed weighted alternatives in terms of both finite sample bias and standard deviation. 

By investigating the theoretical guarantee of the asymptotic variance, we aim to provide a deeper understanding of the statistical properties and performance limits of our proposed estimator. This analysis will shed light on the precision and reliability of our estimator as sample sizes grow and allow for valuable insights into its practical applications.
\subsection{Uncertainty Quantification and Cramer Rao Lower bound}

Since our joint estimator $\hat\theta$ minimizes the negative log-likelihood (2.3), according to Taylor Expansion, we know 
\begin{align*}
	0=\nabla \mathcal{L}(\hat\theta)=\nabla \mathcal{L}(\theta^*)+\nabla^2\mathcal{L}(\theta^*)(\hat\theta-\theta^*)+\textrm{ higher order terms}
\end{align*}
where $\theta^*=(p_0,q_0,\lambda).$
By direct calculation, we obtain
\begin{align*}
	\hat\theta-\theta^*= \frac{-\nabla\mathcal{L}(\theta^*) }{\nabla^2\mathcal{L}(\theta^*)}+o_p(1).
\end{align*}

All involved distributions are binary distributions with bounded support. Moreover, the density function is also continuous to $\theta=(p_0,q_0,\lambda).$ Therefore, the regularity condition holds and $\textrm{Var}(\nabla \ell(\theta^*))=\mathbb{E}[\nabla^2 \ell(\theta^*)]$ (since we are studying the negative log-likelihood). Therefore, we have the following theorem for the asymptotic distribution of $\hat\theta.$
\begin{theorem}\label{asymptotic}
	We have $\hat\theta-\theta^*\rightarrow N(0,I^{-1}(\theta^*))$ where
	$I(\theta^*)=\mathbb{E}[\nabla^2 \ell(\theta^*)]$ is the Fisher information matrix evaluated under $\theta^*.$
\end{theorem}
Based on the presented theorem, we can deduce that our joint estimator has the capability to achieve the minimum variance among all consistent estimators. This finding provides a theoretical validation for our previous observation that our estimator exhibits superior efficiency in terms of standard deviation. It confirms that our estimator possesses favorable statistical properties, reinforcing its potential as an optimal choice for parameter estimation in the given context.

Next, we use some simulations to validate the asymptotic variance in the following section.

\section*{Simulation Validation}
In this section, we validate the asymptotic variance of $\hat\lambda_c$ presented in Theorem \ref{asymptotic}.
In this simulation, we employ the same settings as outlined in Section 3. We repeat the experiments 1000 times and record the normalized estimators, which are obtained by dividing the estimators by the corresponding variances presented in Theorem \ref{asymptotic}, for each replication. The resulting normalized estimators are then used to construct histograms, which are presented below.

\begin{figure}[!p]
	\centering
	\caption{Asymptotic distribution of normalized $\hat\lambda_c-\lambda$ under different cases. (a): $\lambda=1.0,n=500$; (b): $\lambda=1.0,n=1000$; (c): $\lambda=0.9,n=500$; (b): $\lambda=0.9,n=1000$;}
	\begin{tabular}{cc}
		\hskip-30pt\includegraphics[width=0.45\textwidth]{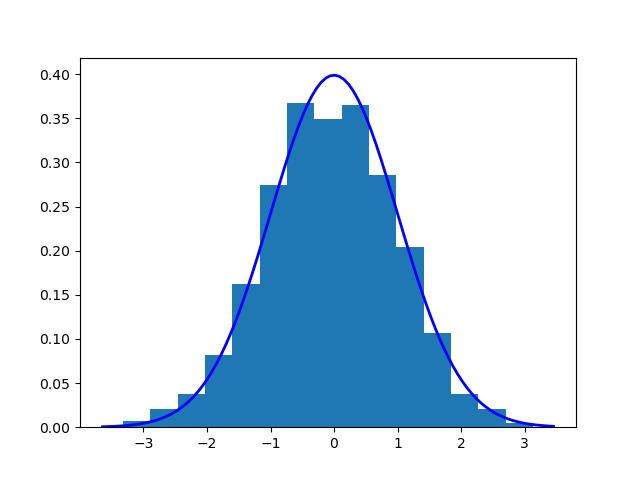}
		&
		\hskip-6pt\includegraphics[width=0.45\textwidth]{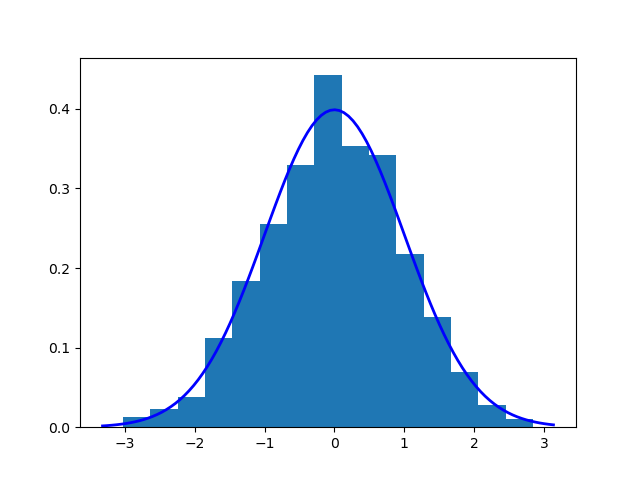}\\
		(a)  & (b) \\
		\hskip-30pt\includegraphics[width=0.45\textwidth]{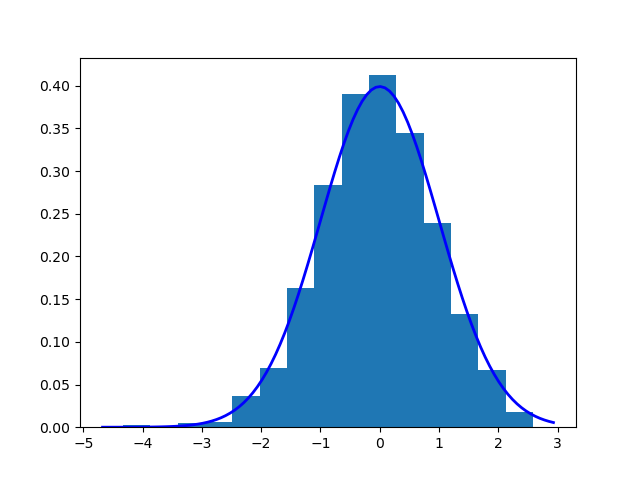}
		&
		\hskip-6pt\includegraphics[width=0.45\textwidth]{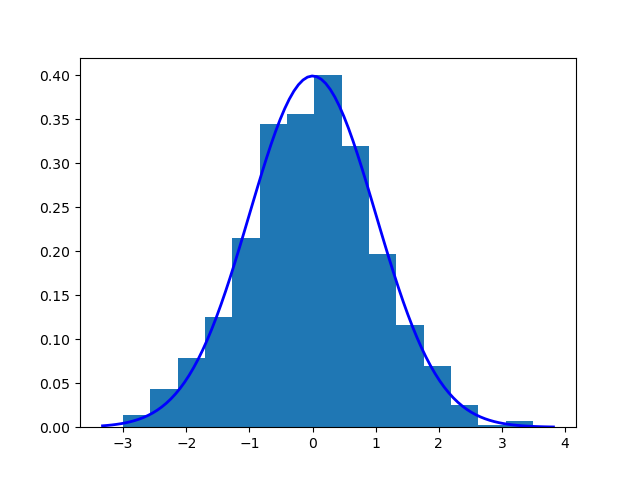}\\
		(c)  & (d) \\
	\end{tabular}
	\label{functions1}
\end{figure}

\begin{figure}[!p]
	\centering
	\caption{Asymptotic distribution of normalized $\hat\lambda_c-\lambda$ under different cases. (e): 
		$\lambda=0.8,n=500$; (f): $\lambda=0.8,n=1000$; (g): $\lambda=0.7,n=500$; (h): $\lambda=0.7,n=1000$;
		(i): $\lambda=0.6,n=500$; (j): $\lambda=0.6,n=1000$;}
	\begin{tabular}{cc}
		\hskip-30pt\includegraphics[width=0.45\textwidth]{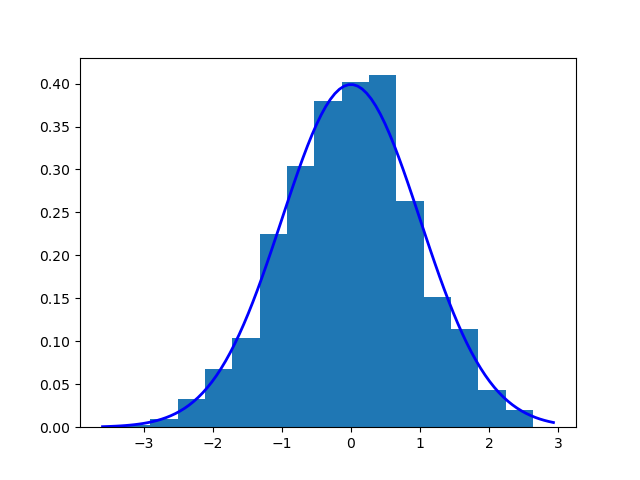}
		&
		\hskip-6pt\includegraphics[width=0.45\textwidth]{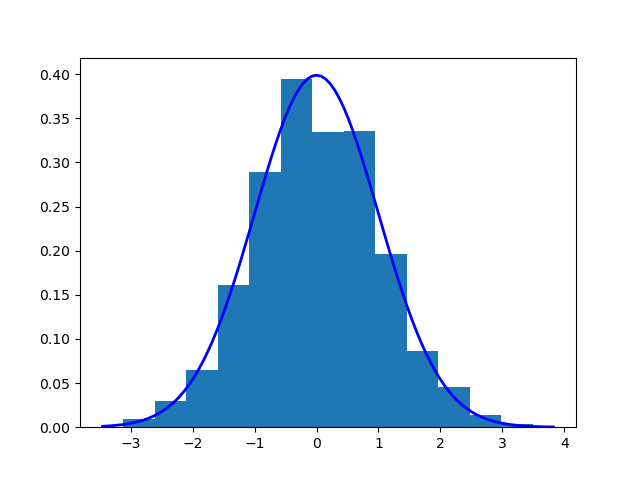}\\
		(e)  & (f) \\
		\hskip-30pt\includegraphics[width=0.45\textwidth]{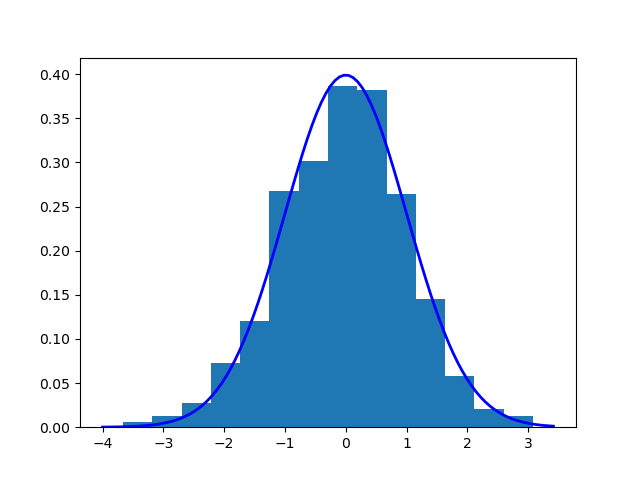}
		&
		\hskip-6pt\includegraphics[width=0.45\textwidth]{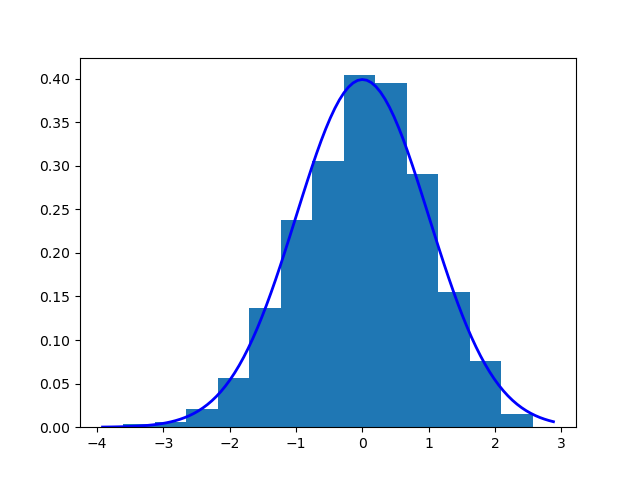}\\
		(g)  & (h) \\
		\hskip-30pt\includegraphics[width=0.45\textwidth]{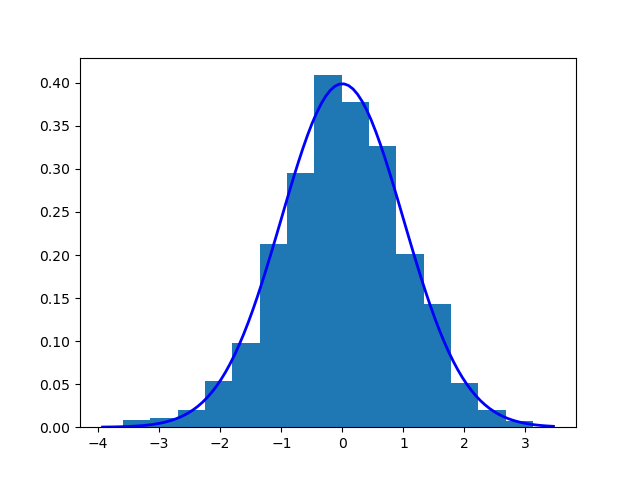}
		&
		\hskip-6pt\includegraphics[width=0.45\textwidth]{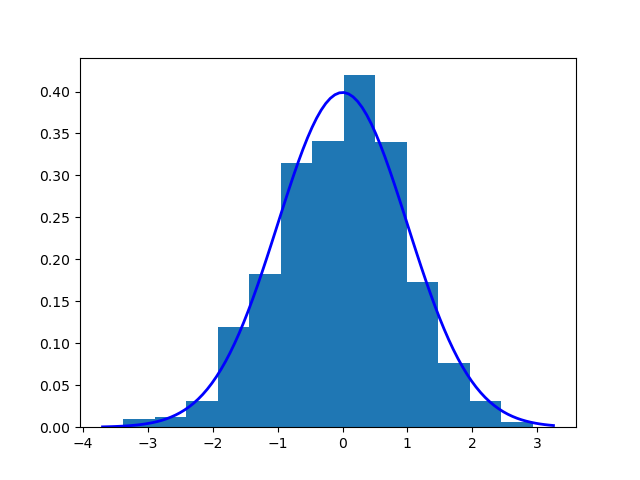}\\
		(i)  & (j) 
	\end{tabular}
	\label{functions2}
\end{figure}

The presented figures consist of two panels in Figures \ref{functions1} and \ref{functions2} , corresponding to experiments conducted with sample sizes of $n=500$ and $n=1000$, respectively. Each figure contains five rows, representing the results obtained for different true parameter values $\lambda={1,0.9,0.8,0.7,0.6}$.

These histograms provide an empirical visualization of the distribution of the normalized estimators, offering insights into the accuracy and precision of our proposed methodology under different true parameter settings and sample sizes.

\bmsection*{Author Biography}

\bibliography{refs}

\end{document}